\RequirePackage[T1]{fontenc}
\documentclass[conference,letterpaper]{IEEEtran}
\IEEEoverridecommandlockouts
\usepackage{cite}
\usepackage{amsmath,amssymb,array,booktabs,graphicx}
\usepackage{url}
\usepackage[hidelinks]{hyperref}
\hypersetup{pdftitle={HyperDBDiag: Multi-Root-Cause Identification and Operator-Level Localization for Database Performance Anomalies},pdfauthor={Hu Wang, Zhengjin Yan, Hong Zhu, Qiong Zuo}}
\newcommand{\indicator}[1]{\mathbf{1}\!\left(#1\right)}
\providecommand{\Description}[1]{}

\makeatletter
\let\arxiv@originaltitlepagestyle\ps@IEEEtitlepagestyle
\def\ps@IEEEtitlepagestyle{%
  \arxiv@originaltitlepagestyle
  \def\@oddfoot{\parbox{\textwidth}{\centering\footnotesize
    This work has been submitted to the IEEE for possible publication. Copyright may be transferred without notice, after which this version may no longer be accessible.}}%
  \let\@evenfoot\@oddfoot}
\makeatother

\begin{document}
\title{HyperDBDiag: Multi-Root-Cause Identification and Operator-Level Localization for Database Performance Anomalies}
\author{\IEEEauthorblockN{Hu Wang, Zhengjin Yan, Hong Zhu, Qiong Zuo\IEEEauthorrefmark{1}\thanks{\IEEEauthorrefmark{1}Corresponding author.}}
\IEEEauthorblockA{Huazhong University of Science and Technology\\
Wuhan, Hubei, China\\
hudawang@hust.edu.cn, d202581876@hust.edu.cn\\
zhuhong@hust.edu.cn, qiong\_zuo@hust.edu.cn}}
\maketitle

\begin{abstract}
Database performance anomalies in modern production environments are not always triggered by a single cause. Instead, they often arise from the concurrent interaction of multiple causes within the same time window, posing substantial challenges to accurate operator-level root-cause analysis. When multiple root causes interact through shared resources, metric anomalies may be amplified, attenuated, or masked. Consequently, salient local anomalies may no longer correspond directly to the actual root causes, making conventional single-path diagnosis and shallow correlation-based methods prone to misdiagnosis.
To address this challenge, we propose HyperDBDiag, a method for multi-root-cause identification and operator-level localization of database performance anomalies. HyperDBDiag connects root causes, performance metrics, SQL templates, and execution operators in a knowledge graph, captures joint anomaly patterns across metrics with a metric hypergraph, and uses a constrained LLM to determine the final root-cause set and provide remediation suggestions. When finer-grained localization is possible, HyperDBDiag also identifies the associated anomalous SQL statements and execution operators.
We evaluate HyperDBDiag on a database performance anomaly dataset constructed using DB-MAGS. Experimental results show that HyperDBDiag effectively reduces graph complexity while improving root-cause localization accuracy.
\end{abstract}

\section{Introduction}\label{sec:introduction}

In real-world database operations, a performance degradation alert often cannot be attributed to a single cause. Instead, the alert may reflect the concurrent effects of multiple root-cause mechanisms within the same time window. Consider a concurrent workload in an online order processing system. An order-detail aggregation query needs to compute statistics over qualifying records in the large \texttt{order\_\allowbreak line} table, but the filtering column \texttt{ol\_\allowbreak quantity} is not indexed, resulting in a full table scan. Meanwhile, a background batch-maintenance transaction is updating the \texttt{orders} table while holding record locks for an extended period. Under such a workload, the full scan of this large table and record-lock waits occur simultaneously, resulting in a mixed anomaly that cannot be adequately explained by a single cause. If a diagnostic system still selects only the single candidate that best matches the locally observed anomaly features, the system may incorrectly attribute the anomaly to sorting-related overhead (\textit{Order By Spill}) because the corresponding pattern closely matches the current observation, thereby failing to identify the true root-cause set $\{\texttt{missing\_\allowbreak index}, \texttt{record\_\allowbreak lock}\}$. Therefore, the key challenge in diagnosing mixed anomalies is not simply to return several correlated candidates, but to identify the actual set of root causes that jointly explains the observed anomaly. Database anomaly benchmarks such as DB-MAGS \cite{Shen2024DBMAGS} and DBPA \cite{Huang2023DBPA} also demonstrate that database performance anomalies may arise from combinations of multiple anomalies through either concurrent or causal interactions. Under such conditions, assuming that each performance anomaly can be primarily explained by a single root cause fails to capture the joint effects of multiple root causes on observed metrics and SQL execution.

Existing approaches have made substantial progress in identifying anomalous metrics, classifying anomaly types, screening suspicious SQL queries, and localizing problematic execution operators. By exploiting KPIs, logs, system calls, or query execution information, these methods progressively narrow the diagnostic scope and provide useful clues from system-level states down to individual queries or operators for single-anomaly diagnosis \cite{Gao2025KPI,Zhang2025ELog,Yan2025Siamese}. However, accurately localizing multiple root causes in mixed-anomaly scenarios remains challenging.

\begin{figure*}[t]
    \centering
    \includegraphics[width=0.64\textwidth]{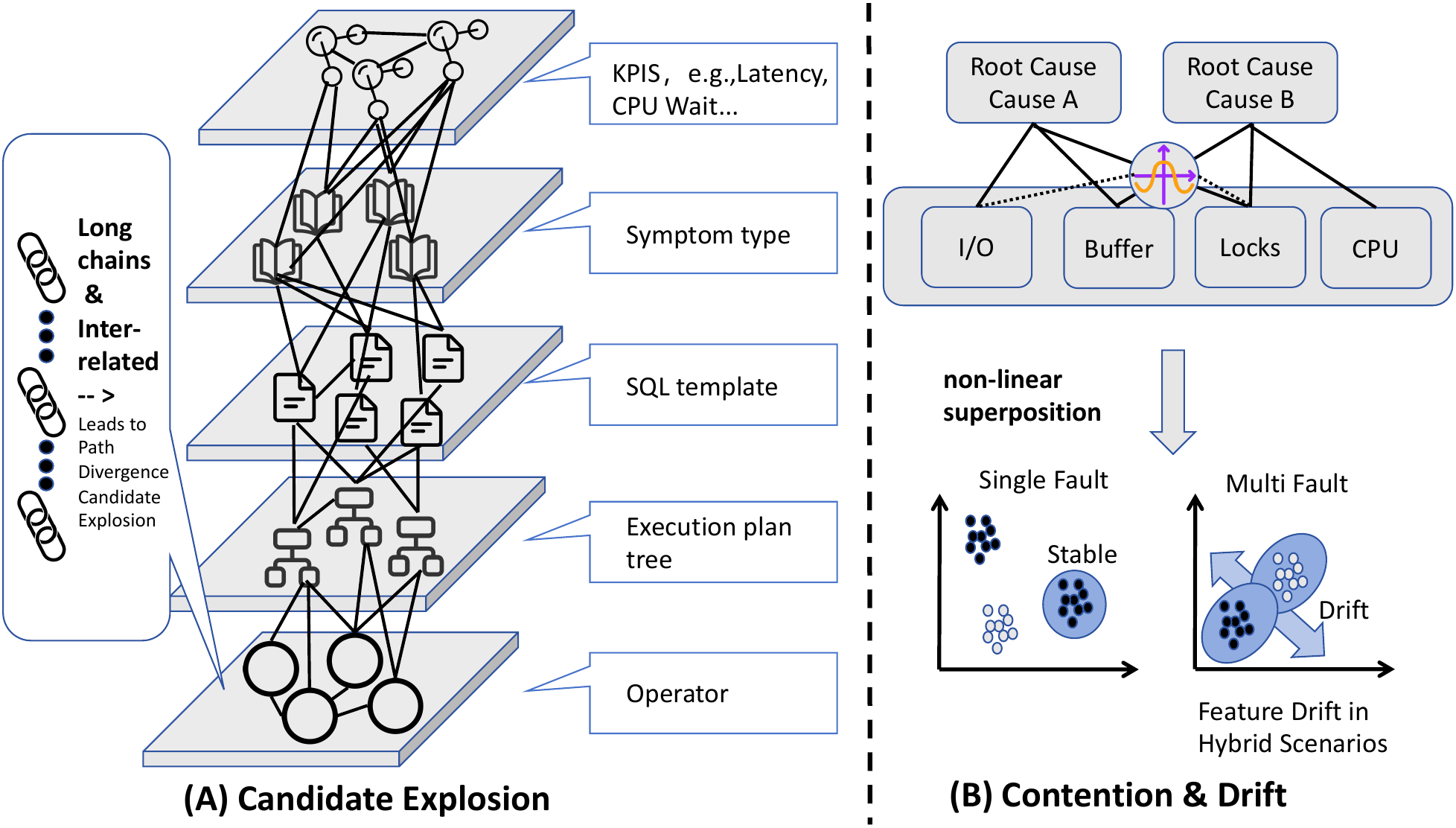}
    \caption{Challenges in mixed-root-cause localization.}
    \Description{Two-part illustration of mixed-root-cause diagnosis challenges: a layered diagnostic graph produces long, interrelated paths and rapid candidate growth, while two root causes jointly affect shared I/O, buffer, lock, and CPU metrics and distort their individual anomaly patterns.}
    \label{fig:architecture}
\end{figure*}

This problem poses two major challenges. First, as illustrated in Figure~\ref{fig:architecture}(a), multiple root causes can cause both the candidate space and the associated evidence chains to grow rapidly. Conventional graph-based approaches typically use binary edges to connect objects such as performance metrics, symptom types, SQL templates, execution plans, and execution operators. Under this representation, a group of metrics that change jointly within the same time window is decomposed and processed independently. During layer-by-layer candidate expansion, one metric may be associated with multiple anomaly symptoms, each symptom may correspond to multiple SQL queries, and each SQL query may further contain an execution plan and a set of execution operators associated with the query. Branches introduced at each layer therefore continue to expand through subsequent layers. Moreover, when different root causes share certain metrics or execution phenomena, these branches may intersect, producing a large number of seemingly plausible candidate paths. In practice, many such paths either repeatedly rely on the same evidence or are valid only for specific combinations of root causes, further enlarging the candidate space and complicating the evidence structure. Second, as shown in Figure~\ref{fig:architecture}(b), concurrent root causes can alter the otherwise relatively clear correspondence between performance metrics and individual root causes. When a single root cause occurs, a subset of metrics can often provide a relatively direct indication of the corresponding anomaly. When multiple root causes occur simultaneously, however, they jointly affect these metrics through shared resources such as I/O, locks, and buffer pools, causing both the magnitude and the combination of metric changes to deviate from the patterns observed under individual root causes. Applying single-root-cause patterns independently in such cases makes the diagnosis susceptible to locally dominant signals, which may result in missed root causes or incorrect attribution.

To address these challenges, we propose HyperDBDiag, a multi-root-cause identification and operator-level localization approach for database performance anomalies.\par
To represent the complex relationships among different diagnostic entities, HyperDBDiag first constructs a knowledge graph that connects the root causes of performance anomalies, performance metrics, SQL templates, and execution operators, allowing candidate root causes to be traced back to their corresponding metric evidence through graph paths. Building on the knowledge graph, HyperDBDiag constructs the Root-Cause-Aware Metric Hypergraph (RCA-MH), which connects multiple performance metrics that change within the same anomaly window through a hyperedge to model their joint anomaly pattern \cite{Antelmi2024HypergraphSurvey,Feng2019HNN,Huang2025HypergraphAnomaly}. RCA-MH then combines this pattern with cross-layer evidence encoded in the knowledge graph to generate candidate root-cause sets. By explicitly modeling groups of jointly changing metrics, the hypergraph representation reduces redundant diagnostic paths associated with specific root-cause combinations and mitigates the risk of misdiagnosis caused by changes in metric-to-root-cause associations.

Motivated by prior applications of large language models (LLMs) to failure analysis and diagnostic query recommendation \cite{Jiang2024Xpert}, HyperDBDiag introduces Evidence-Constrained Set Arbitration (ECSA) following candidate generation. To reduce the risk of missing true root causes, the candidate-generation stage retains multiple plausible root causes. A local structured discriminator organizes each candidate and its associated metric, SQL, and operator evidence from the knowledge graph into an evidence card, determines whether the candidate can be retained as an independent root cause, and identifies candidates that redundantly explain the same phenomenon. Subsequently, an LLM-based arbiter considers the candidate set as a whole, compares the complementary and redundant explanations of different candidates, determines the final root-cause set, and produces corresponding remediation recommendations. When a root cause can be further localized to SQL execution, HyperDBDiag follows the corresponding paths in the knowledge graph to identify the relevant anomalous SQL statements and execution operators \cite{Huang2025OpDiag}.

The main contributions of this work are summarized as follows:
\begin{itemize}
    \item We propose HyperDBDiag, a multi-root-cause identification and operator-level localization approach for database performance anomalies. To the best of our knowledge, HyperDBDiag is the first database performance anomaly diagnosis method to combine complete root-cause set identification with the localization of SQL queries and execution operators.

    \item We design a Root-Cause-Aware Metric Hypergraph (RCA-MH) to model joint metric relationships under multi-root-cause performance anomalies. RCA-MH not only captures patterns of multiple performance metrics changing jointly within the same anomaly window, but also incorporates cross-layer relationships among root causes, SQL templates, and execution operators, providing a comprehensive evidence basis for candidate root-cause generation in mixed-anomaly scenarios.

    \item We propose Evidence-Constrained Set Arbitration (ECSA) for determining the final root-cause set. ECSA combines a local structured discriminator with a constrained LLM-based arbiter. It first determines whether each candidate is supported by independent evidence and then compares the complementary and redundant explanations among candidates. In this way, ECSA retains independently supported root causes that collectively explain the observed anomaly, yielding a more accurate and verifiable root-cause set while also providing actionable remediation recommendations for DBAs.

    \item Experiments on DB-MAGS data show that HyperDBDiag identifies root-cause sets and localizes anomalous SQL queries and execution operators. HyperDBDiag reduces structural relations by 79.90\% and graph traversal time by 79.77\% relative to graph-based methods, and improves root-cause localization accuracy by 45.30 percentage points over the best-performing baseline.
\end{itemize}

\section{Related Work}\label{sec:related-work}

\paragraph{Database Performance Anomaly Diagnosis}
Database diagnosis has progressed from KPI and anomaly-type detection to SQL and operator localization. Early systems, including DBSherlock \cite{Yoon2016DBSherlock}, ADDM \cite{Dias2005Oracle}, CauseRank \cite{Lu2022CausalDiagnosis}, FluxInfer \cite{Liu2020FluxInfer}, and NoSQL \cite{Seo2022NoSQLDiagnosis}, identify abnormal metrics or anomaly types such as missing indexes and lock contention. Sentinel \cite{Glasbergen2020Sentinel} compares fine-grained models of system behavior to identify behavioral differences across workloads and configurations, whereas APOLLO \cite{Jung2019APOLLO} automatically generates and reduces problem-triggering queries to diagnose database performance regressions. iSQUAD \cite{Ma2020SlowQueryDiagnosis}, PinSQL \cite{Liu2022PinSQL}, and RCRank \cite{Ouyang2024RCRank} shift diagnosis toward anomalous SQL by combining query text, execution plans, logs, and KPIs. QueryFormer \cite{Zhao2022QueryFormer} preserves the hierarchy of plan operators, whereas PathSim \cite{Sun2011PathSim}, R-GCN \cite{Schlichtkrull2018RGCN}, HAN \cite{Wang2019HAN}, CloudRCA \cite{Zhang2021CloudRCA}, and KGroot \cite{Wang2024KGroot} exploit structured or heterogeneous dependencies to infer candidate causes. OpDiag \cite{Huang2025OpDiag} further localizes performance anomalies to execution operators. For query-level diagnosis, BALANCE \cite{Chen2023BALANCE} attributes anomalies in target KPIs to candidate SQL queries, DBdoctor \cite{Shi2026DBdoctor} associates fine-grained runtime metrics with individual SQL statements, and Luo et al. \cite{Luo2026RootCauseSQL} further study root-cause SQL prioritization and diagnosis for complex performance issues.

For compound anomalies, DBPA \cite{Huang2023DBPA} and DB-MAGS \cite{Shen2024DBMAGS} provide benchmarks and generation mechanisms covering concurrent or causally related anomalies. Xiang et al. \cite{Xiang2025DistDiagnosis} propose DistDiagnosis, which represents node anomalies and inter-node relationships in distributed databases using a compound-anomaly graph and performs root-cause ranking accordingly. In addition, LogDB \cite{Zhang2025LogDB}, E-Log \cite{Zhang2025ELog}, and SDN \cite{Yan2025Siamese} investigate multivariate logs, elastic logs, and anomaly--normal differences. A survey of cloud-native RCA \cite{Fu2025MicroServiceSurvey} and recent methods based on cross-layer symptom correlation, multimodal causal structures, and temporal causal relationships \cite{Zurkowski2024CloudNativeRCA,Zheng2024MULAN,Lin2024NeuralGranger} further broaden the evidence used for localization. Nevertheless, fine-grained methods generally return a single cause, a dominant attribution path, or a ranked candidate list. They do not explicitly determine which candidates are independent root causes and which are propagated symptoms or redundant explanations. HyperDBDiag addresses this gap by identifying a root-cause set for mixed database anomalies.

\paragraph{Hypergraph Modeling}
Pairwise graphs express joint metric variation through multiple edges or multi-hop propagation. A hypergraph instead connects several metrics in one hyperedge, preserving their co-occurrence directly. Zhou et al. \cite{Zhou2006HypergraphLearning} define normalized propagation using an incidence matrix and vertex- and hyperedge-degree matrices, providing a foundation for information propagation over hypergraphs. Antelmi et al. \cite{Antelmi2024HypergraphSurvey} survey high-order hypergraph representation learning; Silva and Willett \cite{Silva2009HypergraphAnomaly} study hypergraph anomaly detection; and HGNN \cite{Feng2019HNN}, HyperGCN \cite{Yadati2019HyperGCN}, and DHGNN \cite{Jiang2019DynamicHypergraphNN} learn high-order representations. CIE-GAD \cite{Huang2025HypergraphAnomaly} models co-occurrence among adjacent hyperedges, Hyperedge Anomaly Detection \cite{Alam2024HyperedgeAnomaly} detects anomalous hyperedges, Rel-HNN \cite{Alam2025RelHNN} targets relational databases, and CHASE \cite{Zhao2024CHASE} uses causal hypergraphs for multimodal root-cause localization. These studies motivate our metric-hypergraph representation of jointly varying signals; RCA-MH additionally incorporates root-cause evidence to generate candidate sets for database diagnosis.

\paragraph{LLM-Assisted Diagnosis}
LLMs have been applied to collect diagnostic information, interpret operational context, and produce root-cause analyses for database or cloud systems. DB-GPT \cite{Zhou2024DBGPT} discusses the integration of LLMs with database tasks from the perspectives of prompt construction, database-specific fine-tuning, and model design. Xpert \cite{Jiang2024Xpert} recommends queries for failure analysis. D-Bot \cite{Zhou2024DBot}, Panda \cite{Singh2024Panda}, Andromeda \cite{Chen2025RAGConfigDebug}, and DBAIOps \cite{Zhou2026DBAIOps} address report generation, database debugging, configuration diagnosis, and database RCA, respectively. RCACopilot \cite{Chen2024LLMCloudRCA}, OpenRCA \cite{Xu2025OpenRCA}, LocaleXpert \cite{Zhong2026LLMFailureLocalization}, and Ahmed et al. \cite{Ahmed2023CloudIncidentLLM} study cloud incidents, software failures, multimodal microservice failures, and production-incident mitigation.

For RCA, LLM reliability depends on evidence quality and reasoning constraints. The AIOps survey \cite{Zhang2026AIOpsSurvey} and Riddell et al. \cite{Riddell2026LLMReasoningFailure} identify limited evidential support and premature convergence as key concerns. Least-to-Most Prompting \cite{Zhou2023LeastToMost} and Lost in the Middle \cite{Liu2024LostMiddle} show that task decomposition and information organization affect reasoning. LLM-as-a-Judge studies \cite{Gu2026LLMJudgeSurvey,Zheng2023MTBench,Liu2023GEval} likewise emphasize structured inputs, explicit criteria, and standardized outputs. HyperDBDiag therefore organizes candidate causes and evidence locally before asking the LLM to compare candidates within a constrained set, rather than asking it to generate causes directly from raw observations.

\section{System Overview}\label{sec:system-overview}

\begin{figure*}[t]
    \centering
    \includegraphics[
        width=0.72\linewidth,
        keepaspectratio
    ]{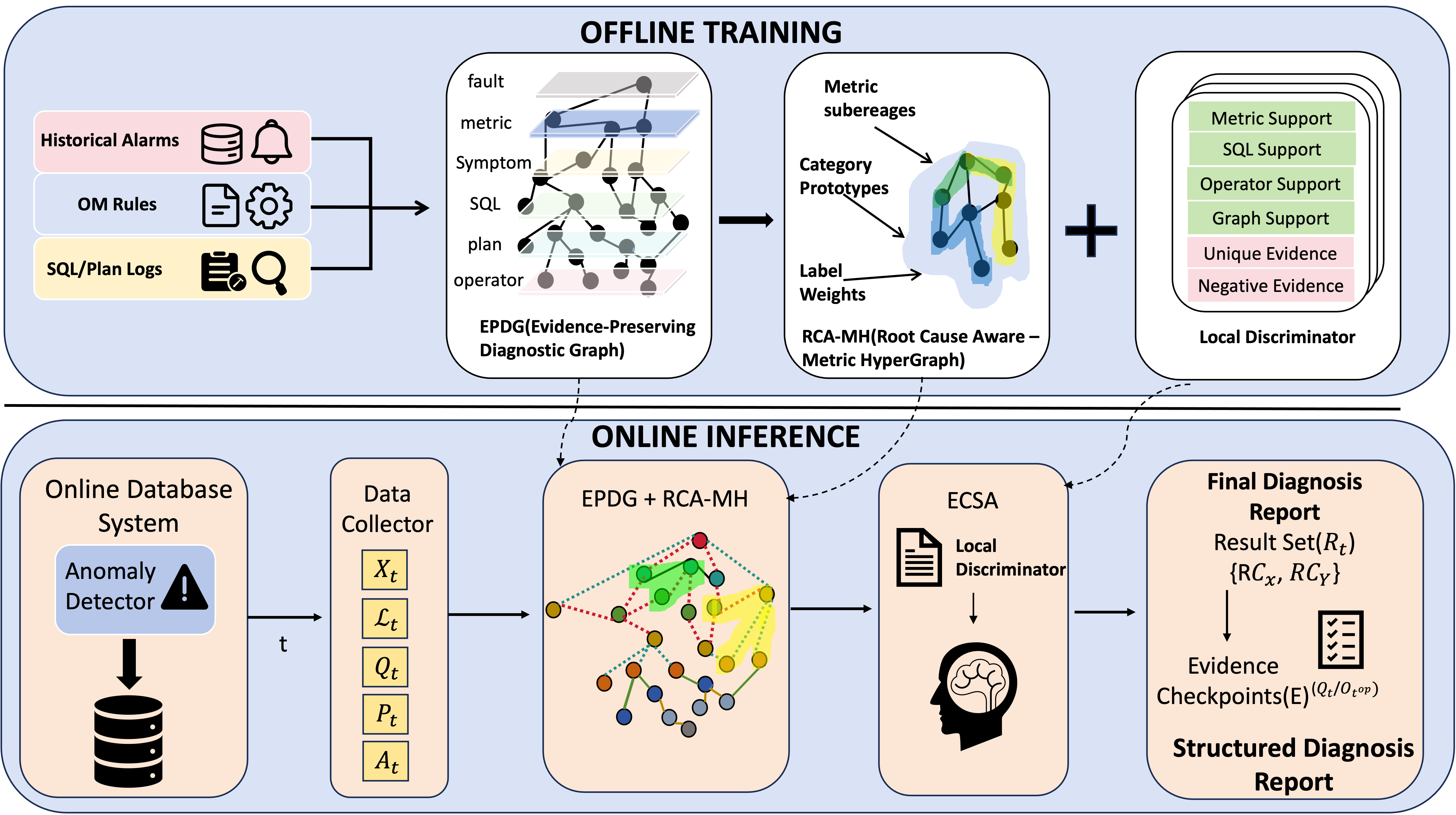}
    \caption{Overview: architecture and workflow of HyperDBDiag.}
    \Description{HyperDBDiag overview with offline construction of EPDG, RCA-MH, and the local discriminator from historical alarms, operational rules, and SQL and plan logs, followed by online data collection, candidate generation, ECSA arbitration, and structured diagnosis reporting.}
    \label{fig:system_overview}
\end{figure*}

This section defines the multi-root-cause diagnosis problem studied in this paper and introduces the overall framework and diagnostic workflow of HyperDBDiag.

\subsection{Problem Definition}\label{sec:problem-definition}

We consider database performance anomaly diagnosis triggered by an anomaly detection module. Our setting covers both anomalies caused by a single root cause and mixed anomalies in which multiple root causes jointly affect the database within the same time window. Suppose that an anomaly detection module detects a performance anomaly at time $t$ and triggers diagnosis. HyperDBDiag uses the operational data collected over the preceding interval of length $\Delta$ as diagnostic evidence. We define the anomaly window as
\begin{equation}
W_t = [t-\Delta, t],
\label{eq:anomaly-window}
\end{equation}
where $t$ denotes the diagnosis trigger time and $\Delta$ is a predefined window length.

In the multi-root-cause setting, different root causes may occur concurrently or may be causally related. To characterize the temporal relationship among multiple root causes within the same diagnostic window, we further divide mixed anomalies into sequential and overlapping types. Let the active intervals of two root causes be $I_A=[s_A,e_A)$ and $I_B=[s_B,e_B)$, respectively. A sequential mixed anomaly occurs when one root cause ends before the other begins, i.e., $e_A \leq s_B$ or $e_B \leq s_A$. Although the root causes do not act simultaneously, the same diagnostic window contains the complete metric changes and execution information associated with both causes, and the diagnostic system must recover the complete root-cause set from that window. An overlapping mixed anomaly occurs when the two active intervals intersect. In this case, the second root cause appears while the first remains active, so the performance metrics and execution states during the overlap are affected by both root causes. Subsequent manifestations induced by a root cause are treated as propagated symptoms rather than additional root causes. Based on the performance metrics, textual clues, and available SQL information collected within $W_t$, HyperDBDiag constructs the diagnostic input as
\begin{equation}
O_t = (x_t, L_t, Q_t, P_t, A_t),
\label{eq:diagnostic-input}
\end{equation}
where $O_t$ denotes the input for the current diagnosis; $x_t$ is the normalized feature vector obtained by preprocessing the performance metrics collected within $W_t$; $L_t$ denotes the clues extracted from relevant textual information and rule-matching results; $Q_t$ is the set of query templates normalized from SQL statements available during execution; $P_t$ denotes the set of available execution plans corresponding to these queries; and $A_t$ denotes the set of execution operators parsed from the available plans. When the corresponding information is unavailable, the sets of query templates, execution plans, and operators may be empty.

Given a predefined root-cause label space $\mathcal{Y}$, the diagnostic task of HyperDBDiag is formulated as
\begin{equation}
\hat{R}_t = F(O_t), \qquad \hat{R}_t \subseteq \mathcal{Y},
\label{eq:diagnostic-task}
\end{equation}
where $F$ denotes the mapping from diagnostic observations to a predicted root-cause set, $\mathcal{Y}$ contains all candidate root-cause labels, and $\hat{R}_t$ is the root-cause set identified by HyperDBDiag. The cardinality of $\hat{R}_t$ is not fixed in advance. Therefore, HyperDBDiag must determine both the number and identities of the root causes, while avoiding treating propagated symptoms or redundant explanations of the same phenomenon as additional root causes.

The diagnosis should also retain supporting evidence for each identified root cause. For root causes that can be further localized to SQL execution, HyperDBDiag also reports the corresponding anomalous SQL statements and key execution operators.

\subsection{HyperDBDiag Framework}\label{sec:framework}

HyperDBDiag consists of three main components: EPDG, RCA-MH, and ECSA. The \emph{Evidence-Preserving Diagnostic Graph} (EPDG) organizes the relationships among performance root causes, performance metrics, and available query templates, execution plans, and execution operators, thereby providing traceable cross-layer evidence for candidate generation.

The \emph{Root-Cause-Aware Metric Hypergraph} (RCA-MH) jointly models multiple metrics that vary within the same anomaly window and combines them with the available cross-layer evidence maintained in EPDG to generate candidate root-cause sets.

Finally, \emph{Evidence-Constrained Set Arbitration} (ECSA) forms evidence cards from candidate root-cause sets and their associated evidence. A local structured discriminator narrows the candidate set, after which a constrained LLM-based arbitrator evaluates whether the remaining candidates complement, cover, or duplicate one another and determines the final root-cause set $\hat{R}_t$.

Figure~\ref{fig:system_overview} summarizes the HyperDBDiag framework, which comprises offline construction and online diagnosis.

\subsection{Offline Construction}\label{sec:offline-construction}

Offline construction uses historical alerts, operational rules, available SQL and execution-plan records, and anomaly samples annotated with root-cause labels. HyperDBDiag builds EPDG to preserve diagnostic paths, trains RCA-MH using metric combinations as hyperedges and root-cause sets as labels, and summarizes the metric variations associated with each root-cause set. These summaries, together with the SQL structures and execution features in EPDG, provide the evidence used by ECSA during online diagnosis.

\subsection{Online Diagnosis}\label{sec:online-diagnosis}

Online diagnosis uses the performance metrics, textual clues, and available SQL information within the anomaly window as diagnostic evidence. RCA-MH uses hypergraph propagation to combine metrics that vary jointly within the same anomaly window and integrates the relationships among root causes, metrics, SQL statements, and execution operators in EPDG to rank candidate root-cause sets. ECSA evaluates how well the metric variations, SQL structures, and execution characteristics associated with each candidate match the current observations, narrows the candidate set, and uses the LLM-based arbitrator to determine the final root-cause set. HyperDBDiag then produces a structured diagnostic report and corresponding remediation suggestions. When fine-grained localization is possible, the report also identifies the anomalous SQL statements and their key execution operators.

\section{Root-Cause-Aware Metric Hypergraph Modeling}\label{sec:rcamh}

\begin{figure*}[t]
    \centering
    \includegraphics[width=0.74\linewidth]{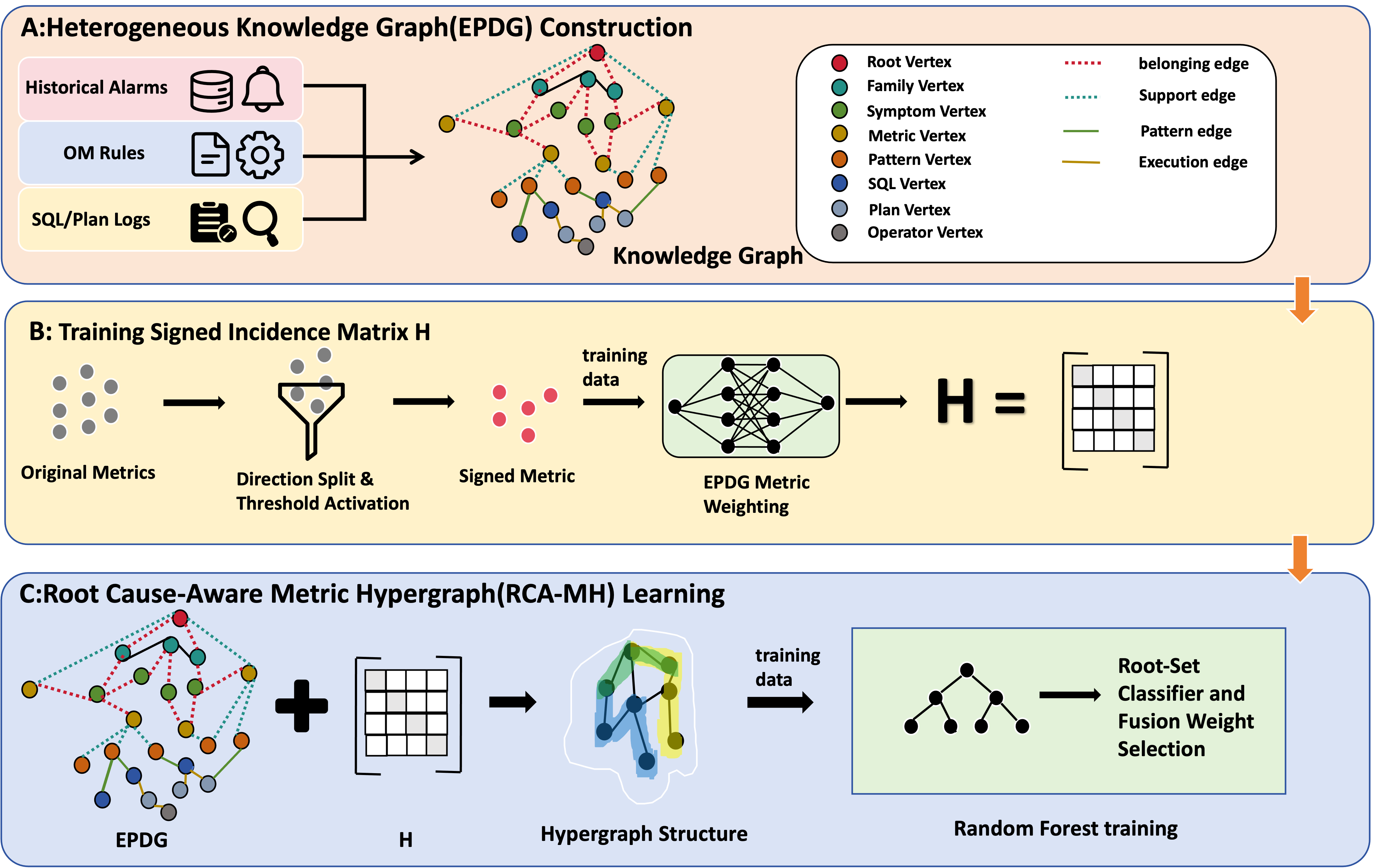}
    \caption{Construction process of the RCA-MH model.}
    \Description{Three-stage RCA-MH modeling process: construction of the heterogeneous EPDG, training of a signed metric incidence matrix from activated metrics and EPDG weights, and learning of the root-cause-aware metric hypergraph and root-set classifier.}
\label{fig:rcamh_modeling}
\end{figure*}

Figure~\ref{fig:rcamh_modeling} illustrates the modeling process of the \emph{Root-Cause-Aware Metric Hypergraph} (RCA-MH) in HyperDBDiag. RCA-MH constructs a metric hypergraph from the training samples and adjusts the training hyperedges using the anomalous metrics associated with individual root causes in EPDG. Accordingly, this section first describes how EPDG organizes diagnostic evidence, then introduces the construction of the metric hypergraph, and finally explains how metric information is propagated through the hypergraph to generate candidate root-cause sets. EPDG organizes the relationships between root causes and performance metrics, together with the currently available SQL structures, execution plans, and operator information, whereas RCA-MH learns the associations between metric combinations observed within the same anomaly window and candidate root-cause sets.

\subsection{EPDG Construction}\label{sec:epdg-construction}

EPDG provides RCA-MH with root-cause--metric associations and retains query-execution information for fine-grained localization. We represent EPDG as a directed heterogeneous diagnostic graph
\begin{equation}
\mathcal{G}_{E} = (\mathcal{V}_{E}, \mathcal{E}_{E}),
\end{equation}
where $\mathcal{V}_{E}$ and $\mathcal{E}_{E}$ denote the node set and edge set, respectively. The EPDG nodes include root causes, root-cause categories, symptoms, metrics, and rule patterns. For root causes that permit fine-grained localization, EPDG additionally contains SQL templates, execution plans, and execution operators. The edge types include hierarchical membership, evidence support, execution relationships, and pattern matching.

EPDG is constructed offline by integrating historical alerts, operational rules, and available SQL and execution-plan logs. Historical alerts provide root-cause labels and their corresponding categories. Operational rules establish explicit associations among root causes, metrics, symptoms, and rule patterns recorded in the rules. The training data further connect each root cause to metrics that consistently exhibit the same direction of change across different collection batches. When SQL logs are available, normalized SQL templates are generated. If execution-plan records are available, the corresponding execution plans and operators are incorporated while preserving the execution relationships among SQL templates, execution plans, and execution operators. When a rule pattern matches a relevant SQL record, an additional pattern-matching edge is introduced. Repeated occurrences of the same object across different records are merged into a single node.

RCA-MH extracts the metric nodes and their connections to root causes from EPDG and uses these relationships to adjust the strengths of the corresponding metrics in the training hyperedges.

\subsection{Metric Hypergraph Construction}\label{sec:metric-hypergraph-construction}

RCA-MH first represents the increase and decrease of each performance metric as two separate directional vertices. RCA-MH then groups the directional vertices whose changes exceed a predefined threshold within the same anomaly window into a single hyperedge. This representation preserves the combination of metrics that change simultaneously during an anomaly while preventing positive and negative changes from canceling each other during propagation. For a training sample, if a metric is connected in EPDG to a root cause in the sample's root-cause set, its incidence strength in the corresponding training hyperedge is increased according to the edge weight, so that metrics more strongly associated with the sample's root causes receive greater weight.

Suppose the offline training set contains $N$ anomalous samples, each consisting of $m$ performance metrics. Let $\mathbf{z}^{(n)}$ denote the normalized metric vector of the $n$-th sample and let $Y^{(n)}$ denote the root-cause set for sample $n$. For the $j$-th metric, we define the magnitudes of its increase and decrease as
\begin{equation}
u_{j,+}^{(n)} = \max\left(z_j^{(n)}, 0\right),
\qquad
u_{j,-}^{(n)} = \max\left(-z_j^{(n)}, 0\right),
\label{eq:directional-magnitudes}
\end{equation}
\begin{samepage}
respectively, and associate them with two directional vertices. The $n$-th sample forms a hyperedge $e^{(n)}$. All samples jointly form an incidence matrix $\mathbf{H}$, in which the incidence strength for direction $s \in \{+,-\}$ is defined as
\begin{equation}
H_{j,s,n}
=
u_{j,s}^{(n)}
\cdot
\indicator{u_{j,s}^{(n)} > \theta}
\cdot
w_j\left(Y^{(n)}\right).
\label{eq:incidence}
\end{equation}
\end{samepage}
Here, $\indicator{\cdot}$ denotes the indicator function, $\theta$ is the metric-selection threshold, and $w_j(Y^{(n)})$ denotes the adjustment weight of the $j$-th metric in the $n$-th training sample. If the metric is connected in EPDG to a root cause of the sample, the metric's weight is increased accordingly. When the sample contains multiple root causes, the corresponding weights of these root causes are averaged. Each column of the incidence matrix corresponds to the hyperedge formed by one anomalous sample. All entries in the matrix are non-negative, while the direction of metric variation is explicitly distinguished by the positive and negative directional vertices.

Each hyperedge directly connects the directional vertices that jointly change beyond the selection threshold within the same anomalous sample. This hyperedge representation therefore avoids constructing pairwise connections among these vertices and reduces redundant propagation paths that could otherwise arise from different combinations of pairwise edges. The ground-truth root-cause sets available during training are used only to adjust the incidence strengths of the corresponding training hyperedges.

\subsection{Hypergraph Learning and Candidate Generation}\label{sec:candidate-generation}

Given the incidence matrix $\mathbf{H}$, RCA-MH adopts the normalized propagation formulation of HGNN~\cite{Feng2019HNN} and aggregates metrics that co-vary within the same anomaly following a ``metric vertex--hyperedge--metric vertex'' propagation scheme. This enables the representation of each metric to incorporate information from other metrics that change together with that metric. The hypergraph propagation operator is defined as
\begin{equation}
\mathbf{P}
=
\mathbf{D}_v^{-\frac{1}{2}}
\mathbf{H}
\mathbf{D}_e^{-1}
\mathbf{H}^{\top}
\mathbf{D}_v^{-\frac{1}{2}},
\label{eq:propagation}
\end{equation}
where $\mathbf{D}_v$ denotes the metric-vertex degree matrix, whose diagonal entry is the sum of the incidence strengths of the corresponding directional vertex over all training hyperedges; $\mathbf{D}_e$ denotes the hyperedge degree matrix, whose diagonal entry is the sum of the incidence strengths of all directional vertices within the corresponding training hyperedge; and $\mathbf{H}^{\top}$ is the transpose of the incidence matrix $\mathbf{H}$. The resulting matrix $\mathbf{P}$ represents the normalized propagation relationships learned from the training hypergraph. RCA-MH uses $\mathbf{P}$ to propagate the current sample's metric representation and uses the difference between the propagated and original representations as the hypergraph-propagation feature.

To use the complete metric combination in each training hyperedge, RCA-MH measures similarities between the current sample and the training hyperedges, aggregates the most similar hyperedges, and uses the difference between this aggregate and the current sample as the hyperedge-matching feature. RCA-MH then concatenates this feature and the hypergraph-propagation feature with the original metric representation and feeds the result into a random forest classifier \cite{Breiman2001RandomForests}, which outputs a probability for each candidate root-cause set.

Candidate root-cause sets can additionally receive supporting evidence from the SQL structural relationships maintained in EPDG. When anonymized SQL structural features can be extracted from the current observation, RCA-MH matches them against the SQL structural relationships associated with root causes in EPDG and aggregates the resulting support for the root causes contained in each candidate set. When such structural features are unavailable, this branch does not affect the ranking of candidate sets.

Let
\begin{equation}
\mathcal{C} = \{C_1, \ldots, C_K\}
\label{eq:candidate-inventory}
\end{equation}
denote the list of candidate root-cause sets, consisting of the distinct root-cause combinations observed during training. For a candidate set $C_k$, let $S_r(C_k)$, $S_h(C_k)$, and $S_e(C_k)$ denote the scores provided by the raw-metric branch, the hypergraph-propagation branch, and the available EPDG structural features, respectively. The final score is defined as
\begin{equation}
\begin{aligned}
S(C_k) ={}& (1-\beta)\bigl[\alpha S_r(C_k) \\
&\quad {}+ (1-\alpha)S_h(C_k)\bigr]
+ \beta S_e(C_k),
\end{aligned}
\label{eq:candidate-score}
\end{equation}
where $\alpha \in [0,1]$ controls the relative contribution of the raw-metric and hypergraph-propagation branches, and $\beta \in [0,1]$ controls the fusion weight of the EPDG structural evidence.

RCA-MH ranks the candidate root-cause sets according to $S(C_k)$ and retains the top-scoring candidates, with the number of retained candidates determined from the training data. RCA-MH outputs these candidate sets together with their scores. As described in the next section, ECSA then determines the final root-cause set based on the evidence available for the current diagnosis.

\section{Evidence-Constrained Set Arbitration}\label{sec:ecsa}

\begin{figure*}[t]
    \centering
    \includegraphics[width=0.84\linewidth]{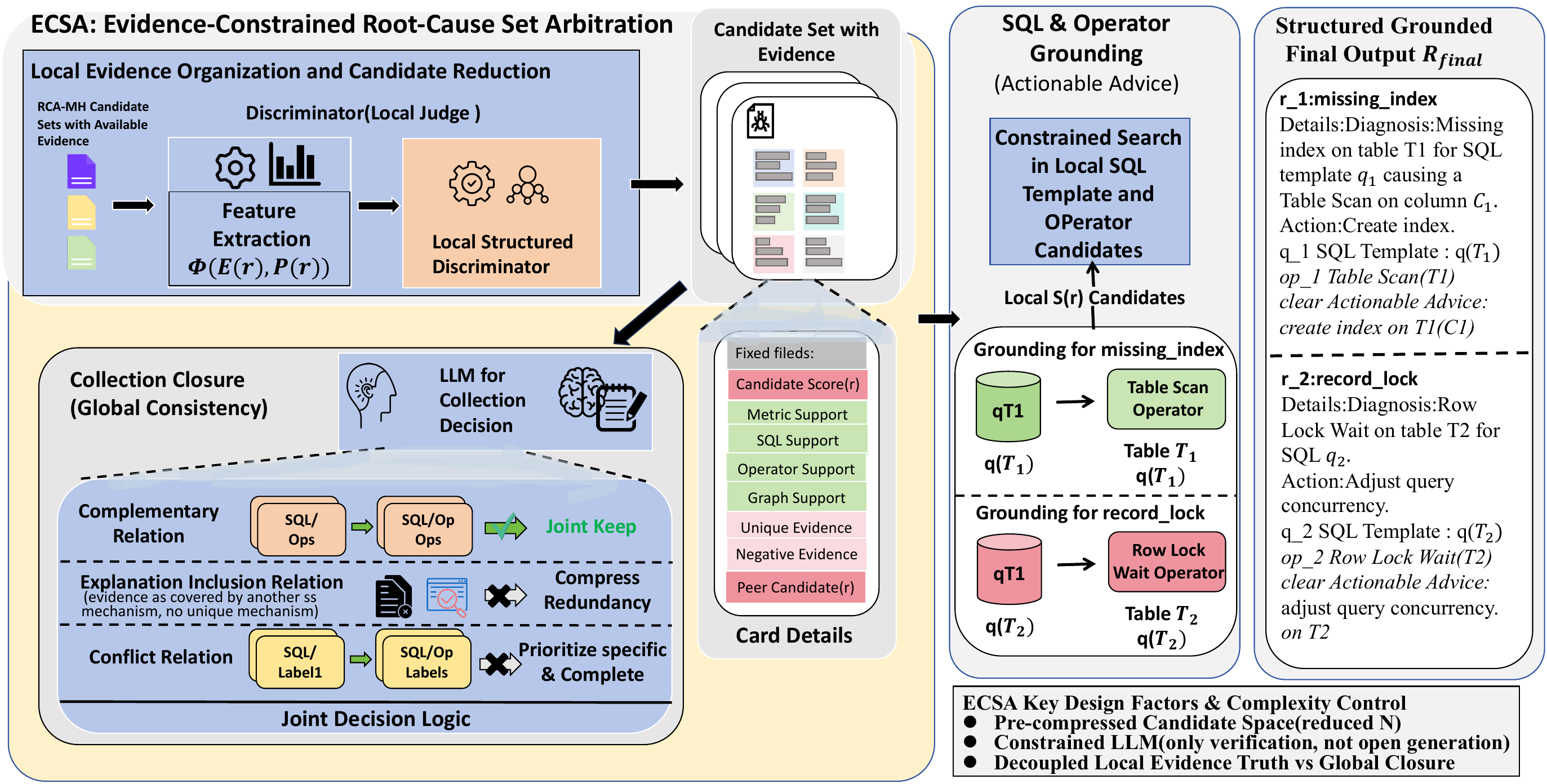}
    \caption{ECSA processing workflow.}
    \Description{ECSA workflow in which a local structured discriminator reduces evidence-backed candidates, an LLM compares complementary, redundant, and conflicting explanations, and constrained SQL and operator grounding produces a structured final diagnosis with remediation advice.}
    \label{fig:ecsa_workflow}
\end{figure*}

This section introduces \emph{Evidence-Constrained Set Arbitration} (ECSA). The previous section produces a ranked list of candidate root-cause sets and retains an evidence summary for each candidate. These candidates may include the true root-cause combination as well as weakly supported or redundant explanations, so they are not directly treated as the final diagnosis.

As illustrated in Figure~\ref{fig:ecsa_workflow}, ECSA takes the candidate root-cause sets and their evidence summaries as input and organizes them into candidate evidence cards. First, the local structured discriminator uses the best-matching candidate set as an initial reference and, among the remaining candidates, retains those supported by current metric observations, SQL structures, or query-execution features. The LLM-based arbitrator then compares whether the retained candidate sets jointly explain the current anomaly and whether any root causes lack independent support or redundantly explain the same phenomenon, and determines the final root-cause set. The local discriminator thus narrows the comparison space, after which the LLM compares the candidate sets using the evidence cards.

\subsection{Candidate Evidence Card Construction}\label{sec:evidence-card}

Given an anomalous observation $\mathcal{O}_t$, the previous section produces a ranked list of candidate root-cause sets. Let
\begin{equation}
\mathcal{C}_K = \{c_1, \ldots, c_K\}
\label{eq:candidate-pool}
\end{equation}
denote the candidate results passed to ECSA, where each $c_i$ represents a candidate root-cause set and $K$ is the number of candidates entering the arbitration stage. ECSA summarizes supporting and inconsistent evidence for each candidate set in a candidate evidence card:
\begin{equation}
\mathbf{a}_i = \left(c_i, E_i^{+}, E_i^{-}\right),
\qquad
c_i \in \mathcal{C}_K.
\label{eq:evidence-card}
\end{equation}
Here, $\mathbf{a}_i$ denotes the evidence card for candidate set $c_i$; $E_i^{+}$ contains evidence supporting the candidate set, including current metric observations consistent with the metric variations associated with the corresponding root-cause set in the training data, as well as SQL-structure or query-execution features without object identifiers observed in the current diagnosis; and $E_i^{-}$ contains evidence inconsistent with the candidate set, including current metrics whose change directions contradict the expected metric variations, as well as the corresponding SQL-structure or query-execution features associated with the candidate root causes but absent from the current observation.

The local structured discriminator first summarizes the metric variations associated with individual root causes and root-cause combinations in the training data and compares the current metrics against these patterns. The discriminator also checks whether the root causes contained in each candidate set receive direct support from the SQL-structure or query-execution features without object identifiers currently available. The resulting evidence cards contain only evidence that supports or is inconsistent with a candidate diagnosis, including features extracted from SQL and execution information, such as full-table reads, predicate type conversions, multi-table joins, sorting, aggregation, and lock waits. The cards exclude raw SQL statements, object identifiers, and complete execution plans, and are used as the input to subsequent local filtering and LLM-based arbitration.

\subsection{Local Candidate Filtering and Set Arbitration}\label{sec:local-arbitration}

After constructing the candidate evidence cards, the local structured discriminator reduces the candidate space by combining metric-pattern consistency with the direct evidence currently available:
\begin{equation}
\mathcal{C}_t^{\mathrm{loc}}
=
J_{\mathrm{loc}}
\left(
\mathcal{O}_t,
\mathcal{C}_K,
\mathcal{P}_{\mathrm{tr}}
\right).
\label{eq:local-filtering}
\end{equation}
Here, $\mathcal{C}_t^{\mathrm{loc}}$ denotes the candidates retained after local discrimination, $J_{\mathrm{loc}}$ denotes the local candidate-filtering procedure, and $\mathcal{P}_{\mathrm{tr}}$ denotes the metric-variation patterns associated with different root-cause sets in the training data. The local structured discriminator only verifies whether the candidates generated by RCA-MH are consistent with the current metrics and direct evidence; the discriminator does not independently predict new root causes.

The LLM-based arbitrator then reads the current anomalous observation together with the evidence cards retained after local filtering and evaluates the candidate root causes jointly by comparing how the candidate explanations complement one another, cover the observed evidence, or duplicate one another:
\begin{equation}
\mathcal{R}_{\mathrm{LLM}}
=
J_{\mathrm{LLM}}
\left(
\mathcal{O}_t,
\left\{
\mathbf{a}_i
\mid
c_i \in \mathcal{C}_t^{\mathrm{loc}}
\right\}
\right).
\label{eq:llm-arbitration}
\end{equation}
Here, $J_{\mathrm{LLM}}$ denotes the constrained LLM-based arbitrator, and $\mathcal{R}_{\mathrm{LLM}}$ denotes the root-cause set selected by the arbitrator. The LLM determines which candidate root causes have independent evidence, which provide complementary explanations, and which represent overlapping or redundant explanations, and accordingly decides which root causes should be retained together.

The arbitration process is constrained by the existing candidates and their evidence, and the LLM determines the diagnosis from the candidates retained by the local discriminator. HyperDBDiag checks whether the result is an existing candidate set and checks the cited evidence; if these checks fail, the local discriminator's result is used to obtain the final root-cause set $\hat{R}_t$.

\subsection{SQL and Execution-Operator Binding}\label{sec:sql-binding}

After the root-cause set has been determined, HyperDBDiag localizes each root cause that can be traced to SQL execution. The localization stage ranks the available SQL queries in the anomaly window and then ranks the operators in the corresponding execution plans. This stage uses the raw SQL statements, object identifiers, and execution plans retained by the system to associate the identified root causes with specific SQL queries and operators. When multiple root causes are identified, HyperDBDiag performs this localization separately for each root cause and attaches the resulting SQL queries and operators as supporting evidence for the corresponding causes, preserving the root-cause--SQL--operator associations for DBA inspection.

\subsection{Diagnostic Result Generation}\label{sec:diagnostic-result}

After these steps, HyperDBDiag produces a structured diagnostic result containing the final root-cause set and the evidence associated with each root cause. For every identified root cause, the report includes the root-cause label, relevant performance metrics, and a summary of the currently available evidence. When SQL- and operator-level localization is available, HyperDBDiag additionally reports the anomalous SQL queries, tables, and key execution operators.

Based on the final root-cause set and the evidence associated with each root cause, the LLM generates remediation suggestions for subsequent inspection and mitigation by DBAs. The structured report preserves the evidence associated with each root cause, supporting subsequent DBA verification and mitigation.

\section{Experiments}\label{sec:experiments}

We evaluate HyperDBDiag on DB-MAGS and focus on answering the following research questions.

\textbf{RQ1:} Compared with existing database anomaly diagnosis methods, can HyperDBDiag more accurately identify the root-cause set and, when applicable, further localize anomalous SQL queries and execution operators?

\textbf{RQ2:} When representing the same metric combinations, can the metric hypergraph reduce the graph size and traversal time?

\textbf{RQ3:} How do the conventional graph, metric hypergraph, EPDG, local structured discriminator, and LLM-based arbitrator individually affect diagnostic performance?

\subsection{Experimental Setup}\label{sec:experimental-setup}

\paragraph{Environment}
We implement HyperDBDiag and all baseline methods in Python, using PyTorch and scikit-learn for model training. All experiments are conducted on an Apple Silicon workstation. HyperDBDiag uses GPT-5.5 for LLM-based arbitration. The filtering thresholds of the local structured discriminator are determined using the training data. We select $\alpha$ and $\beta$ using grouped validation on the training data, with candidate-generation accuracy as the criterion. We determine $K$ from the coverage of the correct root-cause combinations in the training candidate pools and fix $\theta$ at 1.

\paragraph{Dataset}
Following the anomaly definitions in DB-MAGS~\cite{Shen2024DBMAGS}, we reproduce database performance anomalies under a TPC-C workload on MySQL and collect time-series performance metrics and runtime SQL structural evidence before and after anomaly injection. The experiments cover root causes including record locking, missing indexes, implicit type conversion, excessive table joins, sorting, grouping, and full-table queries, spanning three representative categories of database performance problems: lock contention, abnormal access paths, and anomalous query structures.

To construct multi-root-cause scenarios, we combine different anomaly mechanisms in workloads involving the TPC-C \texttt{CUSTOMER} and \texttt{ORDERS} tables. Each combined scenario uses either a missing index or record locking as one root cause and pairs that root cause with implicit type conversion, excessive table joins, sorting, grouping, or full-table queries. We additionally include scenarios in which missing indexes and record locking occur together. For each pair of root causes, the experiments cover single-root and mixed-root scenarios. A single-root scenario injects only one root cause to evaluate diagnosis when no other root cause affects the database concurrently. Mixed-root scenarios include sequential and overlapping cases. In a sequential case, one root cause begins after the preceding root cause has ended; in an overlapping case, the later root cause begins while the preceding root cause remains active, causing their active intervals to partially overlap.

For each injection, we use the anomaly onset as the temporal reference point and collect metrics related to data locks, lock waits, current row-lock waits, running threads, and active sessions. We also extract SQL structural features without object identifiers from runtime statement histories and process information.

For root causes that induce identifiable problematic operators in SQL execution plans, we additionally collect SQL--operator localization data. This part covers missing indexes, implicit type conversion, excessive table joins, sorting, grouping, and full-table queries. For each root-cause type, we design multiple SQL templates and generate different execution plans by varying predicate forms, data-access paths, join structures, sorting columns, and aggregation expressions. The candidate pool contains anomalous SQL queries, SQL queries associated with other root causes, and normal queries with similar structures or execution costs. The diagnostic method is required to identify, from these candidates, the anomalous SQL query corresponding to the current root cause.

\paragraph{Baselines}
We select OpDiag~\cite{Huang2025OpDiag} and DBAIOps~\cite{Zhou2026DBAIOps} as the primary baselines. Both methods are retrained using the same DB-MAGS training data. To let the baselines return root-cause sets, we train a root-count estimator on the training data and select the corresponding number of top-ranked root causes from the scores produced by each method.

The ablation study includes the following variants: a conventional pairwise graph, a metric hypergraph without EPDG, an EPDG-enhanced metric hypergraph, EPDG combined with the local structured discriminator, the metric hypergraph combined with the LLM-based arbitrator, and the complete HyperDBDiag framework.

\paragraph{Evaluation Metrics}
Let $G_i$ and $P_i$ denote the ground-truth and predicted root-cause sets for the $i$-th sample, respectively, and let $N$ denote the total number of samples. We use the following set-level metrics:
\begin{equation}
\mathrm{Exact}
=
\frac{1}{N}
\sum_{i=1}^{N}
\indicator{P_i = G_i},
\end{equation}

\begin{equation}
\begin{aligned}
\mathrm{Precision} &=
\frac{
\sum_{i=1}^{N} |P_i \cap G_i|
}{
\sum_{i=1}^{N} |P_i|
},
\\
\mathrm{Recall} &=
\frac{
\sum_{i=1}^{N} |P_i \cap G_i|
}{
\sum_{i=1}^{N} |G_i|
},
\end{aligned}
\end{equation}

\begin{equation}
\mathrm{F1}
=
\frac{
2 \cdot \mathrm{Precision} \cdot \mathrm{Recall}
}{
\mathrm{Precision} + \mathrm{Recall}
}.
\end{equation}

\emph{Exact} measures the fraction of samples for which the predicted root-cause set exactly matches the ground-truth set. \emph{Precision}, \emph{Recall}, and \emph{F1} measure the identification of individual root-cause components.

For root causes that support further SQL- and operator-level localization, we additionally report \emph{SQL Hit@1} and \emph{Operator Hit@1}, which measure whether the ground-truth SQL query and execution operator, respectively, are ranked first.

\subsection{Diagnostic Performance of Different Methods}\label{sec:diagnostic-performance}

This subsection answers RQ1. Table~\ref{tab:diagnostic_performance} compares the diagnostic performance of OpDiag, DBAIOps, and HyperDBDiag on the DB-MAGS dataset.

OpDiag makes decisions with one classifier per root cause. When multiple root causes induce similar metric changes, different classifiers may assign high scores simultaneously, but these scores do not directly indicate whether the corresponding causes are present together. DBAIOps associates metrics with root causes through an experience graph. Shared metric nodes can give different root causes similar graph-propagation results, which may lead to misdiagnoses or missed causes.

HyperDBDiag organizes metrics that change jointly within the same anomaly window into hyperedges, allowing their joint state to participate directly in candidate generation. EPDG further associates these metric combinations with root-cause mechanisms and SQL structural evidence, thereby distinguishing candidates that exhibit similar metric patterns but arise from different execution mechanisms. For root causes such as implicit type conversion and full-table queries, which may both manifest as large-scale data access, SQL execution features such as type conversion in predicates and full-table scans provide additional discriminative information. The local structured discriminator and constrained LLM-based arbitrator then compare the independent evidence and redundant explanations among candidates to determine the final root-cause set.

For fine-grained localization, HyperDBDiag uses the identified root causes to filter candidates along the root-cause--SQL--execution-plan relationships maintained in EPDG, further linking system-level anomaly diagnosis to the corresponding SQL queries and execution operators. DBAIOps does not include this hierarchical localization procedure and therefore does not report the corresponding results.

\begin{table}[t]
\centering
\caption{Diagnostic performance on DB-MAGS.}
\label{tab:diagnostic_performance}
\setlength{\tabcolsep}{3.3pt}
\begin{tabular}{@{}lrrrrrr@{}}
\toprule
\textbf{Method} &
\textbf{Exact} &
\textbf{Prec.} &
\textbf{Recall} &
\textbf{F1} &
\textbf{SQL H@1} &
\textbf{Op. H@1} \\
\midrule
OpDiag
& 42.58
& 66.02
& 64.77
& 65.39
& 50.87
& 75.00 \\

DBAIOps
& 33.33
& 60.94
& 60.42
& 60.68
& --
& -- \\

HyperDBDiag
& \textbf{87.88}
& \textbf{94.81}
& \textbf{93.37}
& \textbf{94.08}
& \textbf{93.75}
& \textbf{91.67} \\
\bottomrule
\end{tabular}
\end{table}

\begin{figure}[t]
    \centering
    \includegraphics[width=\columnwidth,trim=0 20bp 0 5bp,clip]{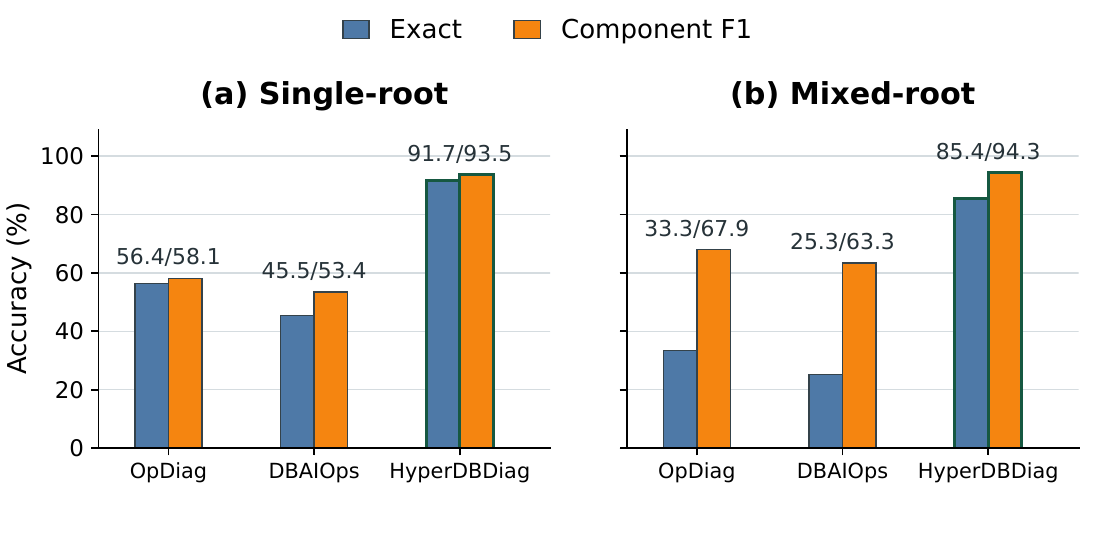}
    \caption{Diagnostic performance under different root-cause cardinalities.}
    \Description{Two side-by-side panels compare Exact and component F1 for OpDiag, DBAIOps, and HyperDBDiag on single-root and mixed-root DB-MAGS cases.}
    \label{fig:cardinality_analysis}
\end{figure}

\paragraph{Single-Root and Mixed-Root Analysis}
To complement RQ1, we divide the DB-MAGS cases into single-root and mixed-root subsets according to their root-cause cardinality. Figure~\ref{fig:cardinality_analysis} reports Exact and component F1 for the two subsets.

HyperDBDiag performs best in both single-root and mixed-root scenarios, with a more pronounced advantage in mixed-root scenarios. Compared with single-root scenarios, mixed-root scenarios require multiple root causes to be identified simultaneously and therefore reveal more clearly the differences among the methods in root-cause-set identification.

\paragraph{Mixed-Root Analysis under Different Temporal Relationships}
To compare diagnosis when multiple root causes occur sequentially or have overlapping active intervals, we divide the mixed-root cases into sequential and overlapping subsets. Table~\ref{tab:occurrence_pattern} reports Exact for the three methods in the two scenarios.

HyperDBDiag performs best in both scenarios. From sequential to overlapping scenarios, HyperDBDiag's Exact decreases only slightly, whereas OpDiag and DBAIOps decline more substantially. During overlap, multiple root causes jointly change the performance metrics. RCA-MH models metrics that change together within the same window as a whole instead of splitting them into independent signals; EPDG uses SQL and execution information to further distinguish causes with similar metric manifestations. Consequently, compared with methods that assess each metric signal independently, HyperDBDiag is less affected by temporal overlap.

\begin{table}[t]
\centering
\caption{Exact for mixed-root scenarios under different temporal relationships.}
\label{tab:occurrence_pattern}
\begin{tabular}{lcc}
\toprule
\textbf{Method} &
\textbf{Sequential} &
\textbf{Overlapping} \\
\midrule
OpDiag
& 39.02
& 21.97 \\

DBAIOps
& 29.17
& 17.42 \\

HyperDBDiag
& \textbf{85.98}
& \textbf{84.09} \\
\bottomrule
\end{tabular}
\end{table}

\paragraph{Fine-Grained SQL and Operator Localization}
Following the hierarchical attribution protocol of OpDiag, we further evaluate the SQL- and operator-level localization stage on the DB-MAGS SQL--operator localization data described above. For each root cause with available SQL evidence, the two methods first rank the blind SQL candidates and then rank the operators in the corresponding execution plan. We report SQL Hit@1, SQL Hit@2, Operator Hit@1, and a joint SQL-Top-2/operator-Top-1 metric, and further compare SQL ranking and joint localization across root-cause cardinalities. The average ranks of the target SQL and target operator provide supplementary analysis. The joint metric requires the target SQL to be ranked within Top-2 and the target operator to be ranked first within that plan. Both methods use the same SQL candidate pool and training/evaluation partition.

\begin{figure}[t]
    \centering
    \includegraphics[width=\columnwidth,trim=0 10bp 0 2bp,clip]{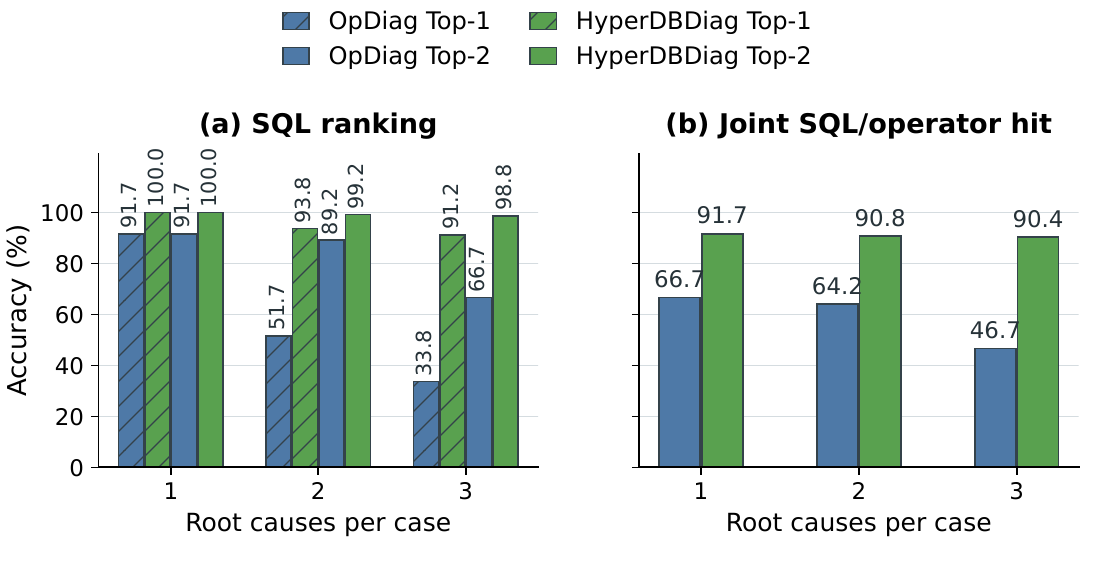}
    \caption{SQL ranking and joint SQL/operator localization under different root-cause cardinalities.}
    \Description{Two panels compare SQL Top-1 and Top-2 accuracy, and joint SQL Top-2 with operator Top-1 accuracy, for OpDiag and HyperDBDiag under one, two, and three concurrent root causes.}
    \label{fig:fine_grained_cardinality}
\end{figure}

Figure~\ref{fig:fine_grained_cardinality} summarizes the results under different root-cause cardinalities. On single-root cases, both methods rank the target SQL reliably. As more concurrent root causes are involved, the SQL ranking accuracy of OpDiag decreases substantially, whereas HyperDBDiag remains close to the performance measured on single-root cases. The joint SQL/operator metric shows the same trend: HyperDBDiag preserves the association between the target root cause, the SQL statement associated with that root cause, and the corresponding execution operator, while OpDiag becomes less reliable as multiple causes compete for shared evidence.

Across all root--SQL pairs, the average target-SQL rank is 2.48 for OpDiag and 1.07 for HyperDBDiag, while the average target-operator ranks are 1.33 and 1.08, respectively. Since the operator rank is computed within the target SQL plan, the joint metric complements the operator ranking by reflecting whether both SQL selection and operator localization are successful.

\subsection{Case Study}\label{sec:case-study}

To further clarify RQ1 and the value of multi-root-cause diagnosis for subsequent DBA investigation and remediation, we select two mixed-anomaly cases from DB-MAGS and then present an SQL/operator-level localization case. The first two cases illustrate the effects of a missed root cause and an incorrect root-cause attribution, respectively. The third case shows how a recovered root-cause set can be connected to concrete queries and execution operators.

\paragraph{Missed diagnosis: missing index and implicit type conversion}
This case is drawn from a DB-MAGS scenario that combines a missing index with an implicit type conversion. A query on the \texttt{customer} table contains an implicit type conversion in the query predicate, while the corresponding access path also lacks the required index. The ground-truth causes are a missing index and an implicit type conversion.
Under the traditional single-root-cause output mode, OpDiag, DBAIOps, and similar methods report one root cause: \texttt{missing\_\allowbreak index}. The implicit type conversion is omitted. A DBA following this result would first inspect the indexes on \texttt{customer}. If the anomaly remains after the index is adjusted, the DBA must reconsider whether the diagnosis was complete, collect the query and execution information again, and continue investigating the unreported type-conversion issue. The diagnosis and remediation process therefore has to be repeated, increasing the investigation time.

HyperDBDiag returns both causes and preserves evidence for the index problem and the predicate conversion. The DBA can inspect the indexes on \texttt{customer} together with the data types and conversion position in the query predicate, reducing the need for repeated diagnosis.

\paragraph{Misdiagnosis: full-table scan and record-lock contention}
In this DB-MAGS case, an unrestricted scan of \texttt{customer} coincides with record-lock contention. The ground-truth root-cause set is \{\texttt{query\_\allowbreak whole\_\allowbreak table}, \texttt{record\_\allowbreak lock}\}. Under the multi-root-cause output adaptation, OpDiag and DBAIOps both return \{\texttt{record\_\allowbreak lock}, \texttt{group\_\allowbreak by}\}, introducing an unsupported grouping cause while omitting the full-table-query cause.

HyperDBDiag correctly returns \{\texttt{query\_\hspace{0pt}whole\_\hspace{0pt}table}, \texttt{record\_\hspace{0pt}lock}\}, supporting inspection of the query's access range and execution path, as well as the lock-holding transaction and lock-wait objects.

\paragraph{SQL and operator localization}
In an operator-level case combining implicit type conversion and a missing index, the ground-truth causes are an implicit type conversion and a missing index. HyperDBDiag localizes the implicit type conversion to \texttt{SELECT COUNT(*)} on \texttt{tpcc10\_\allowbreak test.customer}. The query predicate contains \texttt{c\_\allowbreak w\_\allowbreak id=4 AND c\_\allowbreak d\_\allowbreak id=7} and the disjunctive condition \texttt{c\_\allowbreak last=0 OR c\_\allowbreak last=1}. The key access operator is \texttt{REF} on \texttt{customer}; the execution plan shows that \texttt{c\_\allowbreak last} participates in an implicit numeric comparison, providing direct evidence of the type conversion.

HyperDBDiag simultaneously localizes the missing index to \texttt{SELECT COUNT(*)} on \texttt{tpcc10\_\allowbreak test.order\_\allowbreak line} with the predicate \texttt{ol\_\allowbreak delivery\_\allowbreak d IS NOT NULL}. The execution plan uses access type \texttt{ALL} for \texttt{order\_\allowbreak line}, indicating a full-table scan without an index. OpDiag can localize the SQL query and operator associated with the implicit type conversion, but ranks the \texttt{customer} query first in the missing-index branch and therefore does not form two complete SQL--operator diagnostic chains. DBAIOps remains at the root-cause level and does not identify the corresponding query or operator. The two localization results produced by HyperDBDiag allow the DBA to inspect the type conversion in the \texttt{customer} query and the index and scan path of \texttt{order\_\allowbreak line} separately.

\subsection{Graph Size and Traversal Efficiency}\label{sec:graph-efficiency}

To answer RQ2, we compare the number of relationships and traversal time for pairwise graphs and metric hypergraphs representing the same metric combinations. As shown in Table~\ref{tab:graph_efficiency}, the metric hypergraph requires fewer relationships and less traversal time because each hyperedge preserves a jointly changing metric group without decomposing it into pairwise edges.

\begin{table}[!t]
\centering
\caption{Graph size and traversal efficiency.}
\label{tab:graph_efficiency}
\setlength{\tabcolsep}{3pt}
\begin{tabular}{@{}lrrrr@{}}
\toprule
\textbf{Structure} &
\textbf{Avg. Rel.} &
\begin{tabular}[c]{@{}c@{}}\textbf{Reduction}\\\textbf{(\%)}\end{tabular} &
\begin{tabular}[c]{@{}c@{}}\textbf{Time}\\\textbf{(ms)}\end{tabular} &
\begin{tabular}[c]{@{}c@{}}\textbf{Reduction}\\\textbf{(\%)}\end{tabular} \\
\midrule
Pairwise Graph
& 18,826 & -- & 10.108 & -- \\
Metric Hypergraph
& \textbf{3,783} & \textbf{79.90} & \textbf{2.045} & \textbf{79.77} \\
\bottomrule
\end{tabular}
\end{table}

\subsection{Ablation Study}\label{sec:ablation}

This subsection answers RQ3. Table~\ref{tab:ablation} reports the ablation results of HyperDBDiag on the DB-MAGS dataset. A conventional pairwise graph decomposes metric relationships into multiple binary edges, making it difficult to directly preserve the joint variation pattern of multiple metrics within the same anomaly window. In contrast, the metric hypergraph connects simultaneously activated metrics through a single hyperedge, allowing joint anomaly patterns to participate directly in root-cause-set prediction and thereby making the metric hypergraph better suited to multi-root-cause scenarios.

When only the metric hypergraph is used, the model can capture metric co-occurrence relationships but cannot yet determine which specific root-cause mechanism they correspond to. After EPDG is incorporated, cross-layer relationships among metric combinations, root causes, and SQL structures are used to constrain candidate generation, allowing root causes with similar metric manifestations to be further distinguished by their execution mechanisms.

Using the local structured discriminator alone improves diagnostic performance, indicating that structured evidence helps distinguish among candidates. In the combined variant, the discriminator first filters the candidates and their evidence passed to the LLM, narrowing the arbitration space and reducing input and output token usage. The combined variant outperforms local discrimination alone, indicating that LLM arbitration further improves overall diagnostic performance.

\begin{table}[t]
\centering
\caption{Ablation results of HyperDBDiag on DB-MAGS.}
\label{tab:ablation}
\setlength{\tabcolsep}{3pt}
\begin{tabular}{lcc}
\toprule
\textbf{Method} &
\textbf{Exact (\%)} &
\textbf{F1 (\%)} \\
\midrule
Pairwise Graph
& 28.48
& 64.01 \\

Metric Hypergraph w/o EPDG
& 40.61
& 63.08 \\

EPDG-enhanced Metric Hypergraph
& 57.58
& 74.17 \\

Local Structured Discriminator (w/o LLM)
& 82.73
& 90.12 \\

LLM Arbitration (w/o Local Filtering)
& 87.58
& 93.86 \\

Local Structured Discriminator + LLM Arbitration
& \textbf{87.88}
& \textbf{94.08} \\
\bottomrule
\end{tabular}
\end{table}

\subsection{Parameter Sensitivity}\label{sec:parameter-sensitivity}

\begin{figure}[t]
    \centering
    \includegraphics[width=\columnwidth]{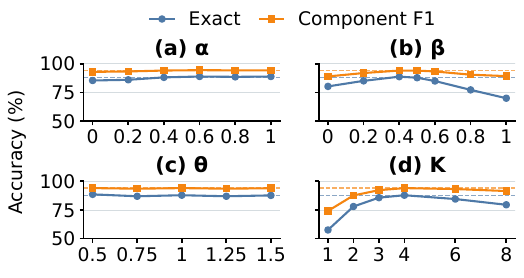}
    \caption{Parameter sensitivity of HyperDBDiag. Dashed lines indicate the results of the configuration used in the main experiments.}
    \Description{Four panels report Exact and component F1 when alpha, beta, theta, or K is varied while other settings remain fixed. Alpha and theta show relatively small changes, while large beta values and candidate counts above four reduce Exact.}
    \label{fig:parameter_sensitivity}
\end{figure}

We vary $\alpha$, $\beta$, $\theta$, and $K$ individually while keeping the other settings fixed, and evaluate the diagnostic performance of the complete HyperDBDiag framework. As shown in Figure~\ref{fig:parameter_sensitivity}, Exact and F1 change relatively little as $\alpha$ and $\theta$ vary. Increasing $\beta$ from a low value initially improves diagnostic performance, but further increases reduce it, indicating that candidate scoring needs to consider both metric information and EPDG structural evidence.

The candidate count $K$ also affects diagnosis. With too few candidates, some correct root-cause combinations cannot reach the subsequent arbitration stage; increasing the candidate count alleviates this problem. In this experiment, $K=4$ achieves the highest Exact. Further increasing $K$ to 6 or 8 leads to more predictions that omit part of the true root-cause set, reducing diagnostic accuracy.

\subsection{Evidence Perturbation Experiments}\label{sec:evidence-perturbation}

To examine how changes in evidence affect diagnosis, we separately give all candidates the same supporting evidence, remove inconsistent evidence, and shuffle the two types of evidence. In another condition, we add misleading information that points to a root cause outside the candidate pool. The candidate sets and other information remain unchanged. We also compare the results obtained by changing the presentation order and repeating diagnosis on the same cases.

As shown in Table~\ref{tab:evidence_perturbation}, the final diagnoses remain among the existing candidates after changes to the evidence. Removing inconsistent evidence reduces Exact and F1, whereas giving candidates identical supporting evidence or shuffling the two types of evidence does not produce a consistent decrease. Changing the presentation order and repeating diagnosis on the same cases also affect some predictions, but do not introduce root causes outside the candidate pool.

\begin{table}[t]
\centering
\caption{Diagnosis results under different evidence conditions.}
\label{tab:evidence_perturbation}
\begin{tabular}{@{}>{\raggedright\arraybackslash}p{0.40\columnwidth}ccc@{}}
\toprule
\textbf{Condition} &
\begin{tabular}[c]{@{}c@{}}\textbf{Exact}\\\textbf{(\%)}\end{tabular} &
\begin{tabular}[c]{@{}c@{}}\textbf{F1}\\\textbf{(\%)}\end{tabular} &
\begin{tabular}[c]{@{}c@{}}\textbf{Diagnoses outside}\\\textbf{the candidate pool}\end{tabular} \\
\midrule
Original evidence & 88.03 & 94.03 & 0 \\
Identical supporting evidence & 88.48 & 94.18 & 0 \\
No inconsistent evidence & 86.97 & 93.37 & 0 \\
Shuffled evidence & 88.79 & 94.34 & 0 \\
Changed order & 87.88 & 94.04 & 0 \\
Out-of-pool information & 88.18 & 94.03 & 0 \\
Repeated diagnosis & 88.03 & 93.84 & 0 \\
\bottomrule
\end{tabular}
\end{table}

\section{Conclusion}\label{sec:conclusion}

This paper presented HyperDBDiag for multi-root-cause identification and operator-level localization of database performance anomalies. EPDG organizes cross-layer relationships, RCA-MH models joint metric changes to generate candidate root-cause sets, and ECSA filters candidates, determines the final set, and produces remediation suggestions. Available execution evidence links diagnosed causes to SQL queries and operators.

Experiments on DB-MAGS-derived data show more accurate complete-set identification than the baselines. The case studies demonstrate the diagnostic value of retaining independently supported causes. The graph comparison shows fewer relationships and lower traversal time, and the ablation study confirms the complementary roles of RCA-MH, EPDG, and ECSA. Future work will examine unseen anomaly combinations, broader workloads, and incomplete SQL or execution-plan evidence.

\section*{Acknowledgment}
This work was supported by Wuhan Dameng Database Co., Ltd. through technical support for fuzz testing targeting database anomaly-handling mechanisms.

\clearpage
\noindent\textit{AI use disclosure.} OpenAI Codex assisted with experimental workflows (Sections IV--VI), plotting (Figures 5--7), and drafting and revising selected text (Sections I, III, V, and VI). GPT-5.5 was used for ECSA arbitration. The authors take responsibility for the manuscript and reported results.

\IEEEtriggeratref{52}
\bibliographystyle{IEEEtran}
\bibliography{sample}

\end{document}